\documentclass[11pt,hyper]{JHEP3}
\usepackage{amsfonts}
\usepackage{amscd}
\usepackage{amssymb}
\usepackage{amsmath}
\usepackage{cite}
\usepackage{latexsym}
\usepackage{mathrsfs}

\renewcommand{\tilde}{\widetilde}

\newcommand{\be}{\begin{equation}}
\newcommand{\ee}{\end{equation}}

\newcommand{\ba}{\begin{eqnarray}}
\newcommand{\ea}{\end{eqnarray}}

\newcommand{\del}{\partial}
\newcommand{\delbar}{\overline{\del}}

\newcommand{\bC}{{\mathbb{C}}}
\newcommand{\bE}{{\mathbb{E}}}

\newcommand{\bR}{{\mathbb{R}}}
\newcommand{\bP}{{\mathbb{CP}}}

\newcommand{\bS}{{\mathbb{S}}}
\newcommand{\bT}{{\mathbb{PT}}}
\newcommand{\bW}{{\mathbb{W}}}

\newcommand{\cA}{{\mathcal{A}}}
\newcommand{\cB}{{\mathcal{B}}}
\newcommand{\cC}{{\mathcal{C}}}
\newcommand{\cF}{{\mathcal{F}}}

\newcommand{\cH}{{\mathcal{H}}}
\newcommand{\cM}{{\mathcal{M}}}
\newcommand{\cN}{{\mathcal{N}}}
\newcommand{\cO}{{\mathcal{O}}}
\newcommand{\cZ}{{\mathcal{Z}}}

\newcommand{\rd}{{\rm d}}
\newcommand{\rD}{{\rm D}}

\newcommand{\e}{{\rm e}}
\newcommand{\im}{{\rm i}}

\title{Supersymmetric Gauge Theories in Twistor Space}
\author{Rutger Boels, Lionel Mason and David Skinner\\
The Mathematical Institute, University of Oxford \\ 24-29 St.
Giles, Oxford OX1 3LP, United Kingdom \\ ${\rm \{boels,
lmason,skinnerd\}@maths.ox.ac.uk}$}

\preprint{5 April, 2006} \keywords{Twistor-string theory, QCD
scattering amplitudes, Twistor theory.}

\abstract{We construct a twistor space action for $\cN=4$ super
Yang-Mills theory and show that it is equivalent to its four
dimensional spacetime counterpart at the level of perturbation
theory. We compare our partition function to the original
twistor-string proposal, showing that although our theory is
closely related to string theory, it is free from conformal
supergravity. We also provide twistor actions for gauge theories
with $\cN<4$ supersymmetry, and show how matter multiplets may be
coupled to the gauge sector.}

\begin{document}

\section{Introduction}

In his construction of twistor-string theory~\cite{Witten:2003nn},
Witten found that the open string sector of the B-model on
$\bP^{3|4}$ coincides with that of $\cN=4$ SYM in spacetime.
States of the open string are described by a (0,1)-form $\cA$ and
expanding this in terms of the fermionic directions of $\bP^{3|4}$
yields component fields which constitute an $\cN=4$ multiplet when
interpreted via the Penrose
transform~\cite{Penrose:1985jw,Ward:1990vs}. Unfortunately,
holomorphic Chern-Simons theory ({\it i.e.} the open string field
theory of the B-model with space-filling
D-branes~\cite{Witten:1992fb})  on $\bP^{3|4}$ only provides the
anti-selfdual couplings of the SYM theory. To overcome this,
building on Nair's observation~\cite{Nair:1988bq} that MHV
amplitudes are supported on holomorphic degree 1 curves in twistor
space, in~\cite{Witten:2003nn} Witten supplemented the twistorial
B-model with D1-instantons, obtaining the missing interactions
from the effective theory of D1-D5 strings. However, his procedure
leads to multi-trace interactions. Such terms would not be present
in connected SYM tree amplitudes, and in~\cite{Berkovits:2004jj}
they were interpreted as mixing in conformal supergravity,
indicating that the D1 instantons prevent the open and closed
string sectors from decoupling.

Despite this fundamental problem, twistor-string theory has
inspired many remarkable and powerful new approaches to Yang-Mills
theory. Following Witten, in~\cite{Roiban:2004vt,Roiban:2004yf}
Roiban, Spradlin \& Volovich were able to `extract' YM tree
amplitudes from the twistor-string by simply discarding the
multi-trace terms. Meanwhile, by considering maximally
disconnected D1 instantons, Cachazo, Svr{\v c}ek \&
Witten~\cite{Cachazo:2004kj} developed tree-level scattering rules
based on using MHV amplitudes with arbitrary numbers of external
particles as primary vertices\footnote{The recursion relations of
Britto, Cachazo \& Feng~\cite{Britto:2004ap}, though even more
succinct, are of less relevance for this paper; they are closer to
the theory in ambitwistor space~\cite{Mason:2005kn}.}. Using
unitarity methods, the MHV diagrams may be tied together to form
loops~\cite{Brandhuber:2004yw} obtaining expressions that agree
with $\cN=4$ 1-loop amplitudes in the
literature~\cite{Bern:1994zx}. In particular, conformal
supergravity does not arise.

We believe that the success of these results strongly indicates
the existence of a theory in twistor space which is exactly
equivalent to spacetime $\cN=4$ SYM. In this paper, we verify this
by presenting a twistor action and showing explicitly that its
partition function coincides with that of the standard spacetime
theory at the perturbative level. A twistor construction and
action for non-supersymmetric Yang-Mills and conformal gravity
appeared in~\cite{Mason:2005zm} together with a formal argument
that attempted to make contact with twistor-string theory---this
latter argument, however, was too formal to be sensitive to the
issue of multi-trace terms.

Our action consists of two parts; a holomorphic Chern-Simons
theory and a term which is closely related, but not identical to
Witten's D-instantons. The action is invariant under the full
group of complex gauge transformations on twistor space, together
with additional twistor cohomological gauge freedom.  This gauge
freedom consists of free functions of six variables rather than
the four variables of spacetime and this extra freedom may either
be partially fixed to cast the theory into standard spacetime
form, or fixed in a way not accessible from space to cast the
theory into a form that makes the MHV diagram formalism
transparent. In a companion paper~\cite{Boels:2006}, we present a
study of perturbation theory based on the MHV form of the action,
showing how it may be used in calculating loop amplitudes. (We
note that the MHV diagram formalism has been derived at tree level
from spacetime considerations
in~\cite{Mansfield:2005yd,Gorsky:2005sf}).

The outline of the paper is as follows. In section 2 we begin by
reviewing the geometry of twistor superspace, and in particular
the reality conditions we employ to descend to Euclidean space.
Similar expositions may be found
in~\cite{Witten:2003nn,Penrose:1985jw,Ward:1990vs,
Ferber:1977qx,Woodhouse:1985id,Harnad:1988rs,Popov:2004rb}, for
example. The main results of the paper are contained in section 3,
where we present our action and show that it is equivalent to
$\cN=4$ SYM at the perturbative level. By breaking the symmetry of
the maximally supersymmetric theory, it is also possible to obtain
twistor actions for YM theories with $\cN<4$. As we discuss in
section 4, this may be done by a method that is similar, but not
identical to working on weighted twistor superspaces. When $\cN<4$
additional multiplets are possible and, following
Ferber~\cite{Ferber:1977qx}, we explain how to construct these and
minimally couple them to the gauge theory. One of the most
important questions our investigation raises is whether, and if so
how, the ideas of this paper are related to string theory. In
section 5 we first refine the arguments of~\cite{Mason:2005zm} to
explain precisely how our twistor action differs from Witten's
original proposal. Nonetheless, we will conclude by proposing that
indeed there are initimate connections with a modification of the
twistor-string.

\vspace{0.5cm}

Our conventions are those of Penrose \&
Rindler~\cite{Penrose:1985jw} (and differ slightly from those of
Witten~\cite{Witten:2003nn}): primed and unprimed capital indices
$A',B',C',\ldots$ and $A,B,C,\ldots$ label elements of $\bS^+$ and
$\bS^-$, the left and right spin bundles, respectively. They are
contracted using the SL(2)-invariant tensors $\epsilon_{AB}$ and
$\epsilon^{A'B'}$, with the conventions
$\omega\cdot\lambda=\omega^A\lambda_A =
\omega^A\lambda^B\epsilon_{BA}$ and
$\pi\cdot\mu=\pi_{A'}\mu^{A'}=\pi_{A'}\mu_{B'}\epsilon^{B'A'}$.
Roman indices $a,b,c,\ldots$ from the beginning of the alphabet
denote elements of the tangent (or cotangent) bundles to
four-dimensional Euclidean space $\bE$. These three sets of
indices can be viewed as `abstract' in the sense that $\rd x^a$ is
a particular 1-form, rather than the $a^{\rm th}$ component of
$\rd x$. This allows us to write $\rd x^a=\rd x^{AA'}$ since the
isomorphism $T^*\bE\simeq \bS^+\otimes \bS^-$ tells us they are
the same geometrical object. If components are required, they may
be obtained using the standard van der Waerden symbols
$\sigma^a_{BB'}$. Greek indices $\alpha,\beta,\gamma,\ldots$
denote elements of the (co-)tangent spaces to $\bC^4$ while Roman
indices $i,j,k,\ldots$ from the
middle of the alphabet 
ranging from $1$ to $\cN$
label the fermionic directions.

\section{The geometry of twistor superspaces}

The projective twistor space of complexified, compactified, flat
spacetime is $\bP^3$. In this paper we will be concerned with the
associated superspaces $\bP^{3|\cN}$ and their relation to
superspacetimes.  To obtain interesting cohomology (and,
physically, to have some notion of an asymptotic region for in and
out states) one must work on the non-compact space $\bT^{3|\cN}$
obtained by removing a $\bP^{1|\cN}$ (corresponding to the
lightcone `at infinity' in spacetime) from $\bP^{3|\cN}$.
$\bP^{3|\cN}$ may be provided with homogeneous coordinates
$[Z^\alpha,\psi^i]=[\omega^A,\pi_{A'},\psi^i]$, defined as always
with respect to the equivalence relation $(Z^\alpha,\psi^i)\sim
(tZ^\alpha,t\psi^i)$ where $t\in\bC^*$. It is then convenient to
remove the $\bP^{1|\cN}$ whose coordinates are
$[\omega^A,0,\psi^i]$.

There are two spaces of immediate interest: the space of
holomorphic lines $\bP^{1|0}\hookrightarrow\bT^{3|\cN}$ and the
space of holomorphic superlines
$\bP^{1|\cN}\hookrightarrow\bT^{3|\cN}$. A holomorphic line is a
$\bP^{1|0}$ linearly embedded in $\bT^{3|\cN}$ as
\begin{equation}
\omega^A=x^{AA'}_-\pi_{A'}\quad\quad\quad
\psi^i=\tilde\theta^{A'i}\pi_{A'},
\label{incid1}
\end{equation}
and hence is parametrized by the 4+2$\cN$ complex coefficients
$(x_-^{AA'},\tilde\theta^{A'i})$. On the other hand, a holomorphic
superline in $\bT^{3|\cN}$ is a $\bP^{1|\cN}$ linearly embedded
via
\begin{equation}
\omega^A=x^{AA'}_+\pi_{A'} - \theta^A_{\,i}\psi^i
\label{incid2}
\end{equation}
and so is parametrized by the 4+2$\cN$ different complex
coefficients $(x_+^{AA'},\theta^A_{\,i})$. It is important to note
that $[\pi_{A'}]$ are the only independent variables on the
$\bP^{1|0}$ of \ref{incid1}, while in \ref{incid2} the independent
variables on the $\bP^{1|\cN}$ are $[\pi_{A'},\psi^i]$. We also
stress that $\theta$ and $\tilde\theta$ are independent complex
fermionic parameters, in particular they are not related by
complex conjugation.

Equations \ref{incid1} and \ref{incid2} show that the $\bP^{1|0}$
intersects the $\bP^{1|\cN}$ whenever
\begin{equation}
\left(x^{AA'}_+-x^{AA'}_--2\theta^A_{\,i}\tilde\theta^{A'i}\right)\pi_{A'}= 0.
\end{equation}
If the contents of the brackets themselves vanish, then this is
true for all $[\pi_{A'}]$ and hence the $\bP^{1|0}$ lies entirely
within the $\bP^{1|4}$. Thus the space $\cM^{4|4\cN}$ of lines
inside superlines inside twistor superspace is complexified
superspacetime, with 4+4$\cN$ complex co-ordinates
$(x^{AA'},\,\theta^A_{\,i},\,\tilde\theta^{A'i})$ where
\begin{equation}\begin{aligned}
x_+^{AA'}&=&x^{AA'} + \theta^A_{\,i}\tilde\theta^{A'i}\\
x_-^{AA'}&=&x^{AA'} - \theta^A_{\,i}\tilde\theta^{A'i}.
\end{aligned}
\label{chmink}
\end{equation}
These equations identify $(x_+,\theta)$ and $(x_-,\tilde\theta)$
as coordinates on complexified chiral and anti-chiral superspace,
respectively.

Throughout most of this paper, we shall be interested in
four-dimensional Euclidean spacetime $\bE$ and its associated
superspaces. In Euclidean signature, complex conjugation sends
primed spinors to primed spinors and unprimed to unprimed by the
formulae
\begin{equation}
\omega^A\rightarrow
\hat\omega^A=(-\overline{\omega^1},\overline{\omega^0})
\qquad\hbox{and}\qquad \pi_{A'}\rightarrow
\hat\pi_{A'}=(-\overline{\pi_{1'}},\overline{ \pi_{0'}})\, .
\label{Eucspin}
\end{equation}
These satisfy $\hat{\hat\omega}^A=-\omega^A$ and
$\hat{\hat\pi}_{A'}=-\pi_{A'}$, so there are no non-vanishing real
spinors. The conjugation induces SU(2) invariant inner products
$|\!|\omega|\!|^2 = \omega^A\hat\omega_A$ and
$|\!|\pi|\!|^2=\pi_{A'}\hat\pi^{A'}$ on $\bS^-$ and $\bS^+$
respectively, and is extended to twistor space in the obvious way
\begin{equation}
Z^\alpha=(\omega^A,\pi_{A'})\rightarrow\hat Z^\alpha = (\hat\omega^A,\hat\pi_{A'})
\label{twconj}
\end{equation}
which again has no fixed points in the projective space. The
conjugation \ref{twconj} provides twistor space with a
non-holomorphic fibration over $S^4$. In the open region $\bT$
this is a fibration over $\bE^4$ given by
\begin{equation}
Z^\alpha=(\omega^A,\pi_{A'})\rightarrow x^{AA'} =
{\omega^A\hat\pi^{A'}-\hat\omega^A\pi^{A'}\over|\!|\pi|\!|^2}.
\label{Efibr}
\end{equation}
It follows from \ref{Eucspin} that $x^{AA'}$ is real, and it is
easy to check that $\det(x^{AA'}) = g(x,x)$ where $g$ is the
standard Euclidean metric on $\bR^4$.

One may extend the conjugation \ref{Eucspin} to the $\cN$
fermionic directions by defining an antiholomorphic map
$s:\bP^{3|\cN}\rightarrow\bP^{3|\overline\cN}$. Our notation
$\overline\cN$ here indicates that $s$ is not an involution.
Generically, $s$ maps the twistor superspace to a different
superspace that has the same body, but whose fermionic directions
are in the complex conjugate representation of the $R$-symmetry
group. If the $R$-symmetry group admits quaternionic
representations ({\it i.e.} when $\cN=2$ or $\cN=4$ and the
$R$-symmetry groups are SU(2) or ${\rm SP}(2)\subset{\rm SU}(4)$),
then $s$ may be involutive on the whole superspace. For example,
when $\cN=4$ one may define $s:\bP^{3|4}\rightarrow\bP^{3|4}$ by
\begin{equation}
s([Z^\alpha,\psi^i])=[\hat Z^\alpha,\hat\psi^{\bar\imath}]
=[\hat\omega^A,\hat\pi_{A'},-\overline{\psi^2},
\overline{\psi^1},-\overline{\psi^4},\overline{\psi^3}].
\label{sact}
\end{equation}
In such cases, one can extend the fibration \ref{Efibr} to the
$\cN=2,4$ superspaces by imposing reality conditions on the
fermions $\tilde\theta^{A'i}$. One finds
$\bT^{3|4}\rightarrow\bE^{4|8}_-$ and
$\bT^{3|2}\rightarrow\bE^{4|4}_-$, where the projection is given
by
\begin{equation}
x_-^{AA'} =
{\omega^A\hat\pi^{A'}-\hat\omega^A\pi^{A'}\over|\!|\pi|\!|^2}
\quad\quad\quad\hbox{and}\quad\quad\quad
\tilde\theta^{A'i}={\psi^i\hat\pi^{A'}-\hat\psi^i\pi^{A'}\over|\!|\pi|\!|^2}
\label{Eac}
\end{equation}
and both $x^{AA'}$ and $\tilde\theta^{A'i}$ are real under $s$.
However, when $\cN=1$ no such reality conditions may be imposed
without forcing the fermions to vanish. (This is a well-known
irritation in discussing Euclidean supersymmetry; see {\it
e.g.}~\cite{vanNieuwenhuizen:1996tv} for a discussion.) Even when
no reality conditions are imposed on the fields, the fibration
\ref{Eac} is still useful as it induces a natural choice of real
contour in the space of $\bP^{1|0}$s.

To make use of all this, in particular the fibration \ref{Efibr}
of the body $\bT$ over $\bE$, it proves convenient to work with
non-holomorphic coordinates $(x_-^{AA'},[\pi_{A'}])$ on $\bT$,
where the scale of $\pi_{A'}$ is projected out. These coordinates
provide a basis for (0,1)-forms which we write as
\begin{equation}
\bar e^0 = {\hat\pi^{A'}\rd\hat\pi_{A'}\over(\pi\cdot\hat\pi)^2}
\quad\quad\quad\quad\bar e^A = {\rd x_-^{AA'}\hat\pi_{A'}\over\pi\cdot\hat\pi},
\label{frame}
\end{equation}
where the factors of $\pi\cdot\hat\pi$ are included for later
convenience and ensure that the basis forms only have holomorphic
weight. The frame is dual to (0,1)-vectors
\begin{equation}
\delbar_0 = (\pi\cdot\hat\pi)\pi_{A'}{\del\phantom{\pi}\over\del\hat\pi_{A'}}
\quad\quad\quad\quad
\delbar_A = \pi^{A'}{\del\phantom{\pi}\over\del x_-^{AA'}}
\label{01}
\end{equation}
in the sense that $\delbar_0\,\lrcorner\,\bar e^0=1$,
$\delbar_A\,\lrcorner\,\bar e^B=\delta_A^{\,B}$,
$\delbar_0\,\lrcorner\,\bar e^A=\delbar_A\,\lrcorner\,\bar e^0=0$
and the $\delbar$-operator is expressed as $\delbar=\bar
e^0\delbar_0+\bar e^A\delbar_A$. Note also that
\begin{equation}
\Omega:={1\over4!}\epsilon_{\alpha\beta\gamma\delta}Z^\alpha\rd
Z^\beta\wedge\rd Z^\gamma\wedge\rd Z^\delta= (\pi\cdot\hat\pi)^4\,
e^0\wedge e^A\wedge e_A \label{emeas}
\end{equation}
where $e^0$ and $e^A$ are the complex conjugates of \ref{frame}.
This basis will be helpful when we relate the twistor and
spacetime SYM actions in the next section using methods that are
described in detail by Woodhouse~\cite{Woodhouse:1985id}.

\section{$\cN=4$ SYM on twistor space}

In this section, we will construct a twistorial action for the
full $\cN$=4 SYM theory. Our theory manifestly contains only
single-trace amplitudes for connected diagrams at tree-level and
is thus free from conformal supergravity. We will see that
different gauge choices allow the twistor action to interpolate
between the usual spacetime $\cN=4$ theory and a Lagrangian
directly adapted to the MHV diagram formalism.

An $\cN=4$ gauge multiplet is CPT self-conjugate and may be
represented on twistor superspace as an element
$\cA\in\Omega^{(0,1)}_{\bT^{3|4}}\left({\rm End}(E)\right)$ where
$\Omega^{(p,q)}_{\bT^{3|4}}$ denotes the space of smooth (not
necessarily holomorphic) $(p,q)$-forms on $\bT^{3|4}$ and
$E\to\bT$ is a vector bundle whose structure group is the
complexification of the spacetime gauge group. We follow
Witten~\cite{Witten:2003nn} in assuming\footnote{These assumptions
amount to taking as a starting point the \v Cech cohomology with
repect to an open cover $\{U_i\}$ of $\bP^{3|4}$, obtained by
pulling back a cover of $\bP^3$ using a fibration
$\bP^{3|4}\rightarrow\bP^3$.  A Dolbeault representative $\cA$ can
be constructed from a \v Cech representative $\cA_{ij}$ defined on
$U_i\cap U_j$ by the formula $\cA=\sum_j \cA_{ij}\bar\del\rho_j$,
where $\rho_i$ is a partition of unity pulled back from $\bP^3$
and subordinate to $U_i$. We leave it to the reader to check that
this has the right properties.} both that $\cA$ has only
holomorphic dependence on $\psi$, and that
$\del/\del\hat\psi^{\bar\imath}\,\lrcorner\,\cA=0$. To fix
notation, the $\psi$-expansion of $\cA$ is taken to be
\begin{equation}
\cA = a + \psi^i\lambda_i + {1\over2!}\psi^i\psi^j\phi_{ij} +
{\epsilon_{ijkl}\over 3!}\psi^i\psi^j\psi^k\chi^l
+{\epsilon_{ijkl}\over 4!}\psi^i\psi^j\psi^k\psi^l b, \label{Aexp}
\end{equation}
where $\{a,\lambda_i,\phi_{ij},\chi^i,b\}$ are smooth (0,1)-forms
on $\bT$ of weight $\{0,-1,-2,-3,-4\}$ respectively.

\subsection{The twistorial $\cN=4$ action}

Our twistor action $S[\cA]$ will be expressed as a sum
$S[\cA]=S_1[\cA]+S_2[\cA]$ as follows. The kinetic terms and
anti-selfdual interactions of $\cN=4$ SYM theory may be described
by a holomorphic Chern-Simons theory on $\bT^{3|4}$ with
action~\cite{Witten:2003nn}
\begin{equation}
S_1[\cA]={\im\over2\pi}\int\Omega\,\rd^4\psi\,\wedge{\rm tr}
\left\{\cA\wedge\delbar\cA+{2\over3}\cA\wedge\cA\wedge\cA\right\}
\label{asd}
\end{equation}
where ${\rm tr}$ indicates a trace using a Killing form on $E$
(and involves a choice of Hermitian metric on the fibres).
$\Omega$ was defined in equation \ref{emeas} and
$\Omega\,\rd^4\psi$ is a holomorphic volume form on the
superspace. The action is invariant under gauge transformations
\begin{equation}
\delbar+\cA \rightarrow g(\delbar+\cA)g^{-1}
\label{gauge}
\end{equation}
where $g$ is an SU($N$)-valued section of $E\rightarrow\bT^{3|4}$
that is homotopic to the identity, and we require $g\rightarrow 1$
asymptotically.  This is a considerably greater freedom than in
spacetime because $g$, like $\cA$, is only required to be smooth
and so is a function of six variables. Ordinarily, one chooses the
numerical coefficient of the Chern-Simons action on a real
3-manifold $M$ to ensure that the partition function is invariant
under gauge transformations $g$ that do not map to the identity in
$\pi_3(G)$. The issue does not arise here because, on twistor
(super)space, $b_3=0$ and the normalization of \ref{asd} is
arbitrary. $S_1$ leads to an equation of motion $\delbar\cA +
[\cA,\cA] =0$ which implies that the (0,2)-component of the
curvature of $\cA$ vanishes. Hence, on-shell $\delbar_\cA$ defines
an integrable complex structure on $E$.  $E$ is then holomorphic,
and so describes an anti-selfdual solution of the $\cN=4$ SYM
equations in complex spacetime via the Penrose-Ward correspondence
(see {\it e.g.}~\cite{Penrose:1985jw,Ward:1990vs}).

To obtain the full $\cN=4$ SYM theory, one considers the action $S = S_1 + S_2$ where
\begin{equation}
S_2[\cA]=-\kappa\int\rd\mu\;\log\det\left((\delbar+\cA)_{L(x_-,\tilde\theta)}\right)
\label{d1}
\end{equation}
where $\kappa$ is a coupling constant (later to be identified with
$g^2_{\rm YM}$). The fibre of $\bT^{3|4}\rightarrow\bE^{4|8}_-$
over a point $(x_-,\tilde\theta)\in\bE^{4|8}_-$ is a $\bP^{1|0}$
that we denote by $L(x_-,\tilde\theta)$. In forming \ref{d1} we
first restrict the Cauchy-Riemann operator $\delbar_\cA$ to
$L(x_-,\tilde\theta)$, then construct the determinant of this
operator and finally integrate the logarithm of this determinant
over the space of lines $\bP^{1|0}\hookrightarrow\bT^{3|4}$ (as
discussed in section 2, this is antichiral superspace) using the
measure
\begin{equation}
\rd\mu:={1\over 4!}\epsilon_{abcd}\,\rd x^a\wedge\rd x^b\wedge\rd
x^c\wedge\rd x^d\;\rd^8\tilde\theta. \label{meas}
\end{equation}
Note that because we are already integrating over all
$\tilde\theta$s, $\rd x_-$ may be equated with $\rd x$.

The determinant of a $\delbar$-operator is not really a function,
but a section of a line bundle over the space of connections as
discussed by Quillen~\cite{Quillen}. The line bundle can be
provided with a metric and a connection, but in general these may
not be flat, and the bundle itself could be non-trivial. However,
this line bundle must be trivial over the space of $\bP^{1|0}$
fibres since this is antichiral Euclidean space $\bE^{4|8}_-$,
which doesn't have sufficient topology. (The line bundle would
similarly be trivial over $S^{4|8}$ but we would have to argue
more carefully for more complicated spacetimes.) We still have the
freedom to choose a trivialisation, but this freedom can be
reduced somewhat by following an observation of Quillen that, by
picking a base-point $\cA_0$ in the space of connections, we can
adjust the metric on the determinant line bundle using the norm of
$\cA-\cA_0$ on the $\bP^1$ so that the associated Chern connection
is flat.  Thus we can trivialize the bundle (up to a constant)
using this flat connection once we have picked a base-point in the
space of connections.  In the associated flat frame we are
justified in treating the determinant na{\"\i}vely as a function
and may integrate its logarithm over $\bE^{4|8}$. However, as is
familiar in physics from anomaly calculations, the dependence of
this trivialization on a fixed background connection $\cA_0$,
which can be taken to be flat, nevertheless breaks gauge
invariance. Indeed, under \ref{gauge} the determinant varies as
\begin{equation}
\det(\delbar+\cA)_L \rightarrow \exp\left({1\over2\pi}\int_L
g^{-1}\del g\wedge\cA\right)\det(\delbar+\cA)_L. \label{detgauge}
\end{equation}
In~\cite{Berkovits:2004jj} the determinant's lack of gauge
invariance was cited as additional evidence for coupling between
the open and closed sectors of the twistor-string, leading to
conformal gravity, as gauge invariance may be restored by a
compensating transformation of the $B$-field in the closed string
sector. Here though, we are concerned with
$\log\det\delbar_\cA|_L$  which under gauge transformations
acquires an additive piece integrated over one copy of the fibre
$L$. Since this term is at most quintic\footnote{It is possible
that $\del g$ introduces a $\tilde\theta$ that is independent of
the combination $\psi^i=\tilde\theta^{A'i}\pi_{A'}$.} in
$\tilde\theta$, it will not survive the Berezinian integration in
$\rd\mu$. Hence the full action is gauge invariant without
recourse to the closed string sector.

The real justification for our action is that it is precisely
equivalent to $\cN=4$ SYM in spacetime, as we shall soon make
clear. However, as a preliminary check notice that $S_1$ and $S_2$
each contain only single-trace interactions (recall that $\log\det
M={\rm tr}\log M$) and that the PGL($4|4,\bC$) transformation
$[Z^\alpha,\psi^i]\mapsto[Z^\alpha,r\psi^i]$ with $r\in\bC^*$
induces the transformations $S_1\mapsto r^{-4}S_1$ and $S_2\mapsto
r^{-8}S_2$ (because $\psi^i\mapsto r\psi^i$ with $\pi_{A'}$
invariant implies $\tilde\theta^{A'i}\mapsto
r\tilde\theta^{A'i})$. Although $S_2$ is non-polynomial in $\cA$,
this scaling together with the induced action $\lambda\mapsto
r^{-1}\lambda$, $\phi\mapsto r^{-2}\phi$, $\chi\mapsto r^{-3}\chi$
and $b\mapsto r^{-4}b$ shows immediately that these component
fields can only appear with certain powers. In particular, $S_2$
is at most quadratic in $b$. Moreover, following Witten's
reasoning, the partition function $\cZ(\hbar,\kappa) = \int{\rm
D}\cA\, \e^{-S[\cA]/\hbar}$ can be made invariant under the
$r$-scaling if we declare that $\hbar\mapsto r^{-4}\hbar$ and
$\kappa\mapsto r^4\kappa$. Conservation of the associated charge
then demands that an $l$-loop contribution to an amplitude with
$n_{\lambda}$, $n_\phi$, $n_\chi$ and $n_b$ external fields of
types $\lambda$, $\phi$, $\chi$ and $b$ respectively must scale
like $\hbar^{l-1}\kappa^d$ where
\begin{equation}
4d = 4n_b + 3n_\chi +2n_\phi +n_\lambda +4(l-1).
\label{scaling}
\end{equation}
In particular, this implies that all amplitudes vanish unless
$3n_\chi+2n_\phi+n_\lambda=4m$ with $m$ a non-negative integer. As
discussed in~\cite{Witten:2003nn}, this is exactly the behaviour
of the spacetime $\cN=4$ action.

\subsection{Equivalence to $\cN=4$ SYM on spacetime}

Let us now validate our claim that $S[\cA]$ is equivalent to
$\cN=4$ SYM on spacetime. To begin, we must partially gauge-fix
$\cA$ to remove the extra symmetry beyond the spacetime gauge
group. To achieve this, expand $\cA$ in the basis \ref{frame} as
$\cA=\bar e^0\cA_0+\bar e^A\cA_A$ and impose the gauge condition
\begin{equation}
\delbar_L^*\cA_0=0
\label{spgauge}
\end{equation}
on all fibres. The notation $\delbar_L$ means the
$\delbar$-operator on the $\bP^1$ fibre labelled by
$L(x_-,\tilde\theta)$ so we are requiring that $\cA_0$ be
fibrewise co-closed with respect to the standard Fubini-Study
metric of each $\bP^1$. Because $\cA$ is holomorphic in $\psi$,
for \ref{spgauge} to hold for all $\tilde\theta$, it must hold for
each component of the $\psi$-expansion separately, so $\delbar_L^*
a_0=\delbar_L^*\lambda_{i0}=\delbar_L^*\phi_{ij\,0}=\delbar_L^*\chi^i_{\,0}=\delbar_L^*
b_0=0$. Since the fields are (0,1)-forms and dim$_{\bC}\,L=1$,
they are all $\delbar$-closed automatically. Also requiring them
to be co-closed ensures that they harmonic along the fibres, so
Hodge's theorem tells us that in this gauge, the restriction of
the component fields to the fibres are End$(E)$-valued elements of
$H^1(\bP^1,\cO(n))$ where $n$ runs from 0 to $-4$. However,
$H^1(\bP^1,\cO)=H^1(\bP^1,\cO(-1))=0$ so that
$a_0=\lambda_{i0}=0$, while the other fields may be put in
standard form~\cite{Woodhouse:1985id}
\begin{equation}
\begin{aligned}
a&=\bar e^Aa_A(x,\pi,\hat\pi) &\lambda_i=&\bar e^A\lambda_{i\,A}(x,\pi,\hat\pi)\\
\phi_{ij}&=\bar e^0 \Phi_{ij}(x)+\bar e^A\phi_{ij\,A}(x,\pi,\hat\pi)
&\chi^i=&2\bar e^0{\tilde\Lambda^i_{\,A'}(x)\hat\pi^{A'}\over\pi\cdot\hat\pi}
+ \bar e^A\chi^i_{\,A}(x,\pi,\hat\pi)\\
b &=3\bar e^0{B_{A'B'}(x)\hat\pi^{A'}\hat\pi^{B'}\over(\pi\cdot\hat\pi)^2}+\bar e^A b_A(x,\pi,\hat\pi)
\end{aligned}
\label{spcompts}
\end{equation}
where, as indicated, $\Phi$, $\tilde\Lambda$ and $B$ depend only
on $x$ and numerical factors are included in the definition of
$\tilde\Lambda$ and $B$ for later convenience. The $\cA_A$
components are as yet unconstrained. The non-trivial step in
achieving this gauge choice is in setting $a_0=0$, which implies
that we have found a frame for $E$ that is holomorphic up the
fibres of $\bT^{3|4}\rightarrow \bE^{3|4}$.  This requires that
the bundle $E$ is trivial up the fibres (a standard assumption of
the Ward construction) which is not guaranteed in general, but, in
a perturbative context will follow from a smallness assumption on
$a_0$.

We have only restricted $\cA$ to be fibrewise harmonic, rather
than harmonic on all of twistor space, so there is some residual
gauge freedom: any gauge transformation by $h$ satisfying
$\delbar^*_L\delbar_L h=0$ for all fibres $L$ leaves \ref{spgauge}
unchanged. If $\delbar^*_L\delbar_L h=0$, $h$ is a globally
defined solution to the Laplacian of weight zero on a $\bP^1$, so
is constant on each fibre by the maximum principle. Hence $h=h(x)$
only and the residual gauge freedom is precisely by spacetime
gauge transformations.

To impose our gauge choice in the path integral, we should include
ghosts $c\in\Omega^{(0,0)}_{\bT^{3|4}}({\rm End}E)$ and antighosts
$\bar c\in\Omega^{(0,3)}_{\bT^{3|4}}({\rm End}E)$ together with a
Nakanishi-Lautrup field $m=[Q,\bar
c]\in\Omega^{(0,3)}_{\bT^{3|4}}({\rm End}E)$ where $Q$ is the BRST
operator. The gauge-fixing term
\begin{equation}
\int\Omega\,\rd^4\psi\wedge{\rm tr}\left[Q, \bar
c\left(\delbar_L^*\cA_0\right)\right]
=\int\Omega\,\rd^4\psi\wedge{\rm
tr}\left\{m\left(\delbar_L^*\cA_0\right)+\bar
c\delbar_L^*\left[\delbar_0+\cA_0,c\right]\right\} \label{gf}
\end{equation}
imposes \ref{spgauge} upon integrating out $m$, whereupon
$\cA_0\sim(\psi)^2$ as we have seen. The ghost kinetic term $\bar
c\delbar^*\delbar c$ only involves terms whose coefficient fields
have net $r$-charge $-4$ in an expansion of $\bar cc$, while the
ghost-matter interaction $\bar c[\cA,c]$ picks out terms of net
$r$-charge $-2$, $-1$ and 0 from the $\bar cc$ expansion.
Considering the $(\bar c,c)$ couplings as a $4\times4$ matrix with
rows and columns labelled by $r$-charge, we see that this matrix
has upper-triangular form, with ghost-matter mixing occuring only
off the diagonal. Hence the Fadeev-Popov determinant is
independent of $\cA$ and the ghost sector decouples from the path
integral.

In terms of the expansion \ref{spcompts}, the Chern-Simons part of
the action becomes
\begin{multline}
S_1[\cA]={\im\over2\pi}
\int{\Omega\wedge\bar\Omega\over(\pi\cdot\hat\pi)^4}\, {\rm
tr}\left\{3{
B_{A'B'}\hat\pi^{A'}\hat\pi^{B'}\over(\pi\cdot\hat\pi)^2}\left(\pi^{C'}{\del
a^A\over\del x^{AC'}} + {1\over2}[a^A,a_A]\right)\right.\\
+2{\tilde\Lambda^i_{A'}\hat\pi^{A'}\over\pi\cdot\hat\pi}\left(\pi^{B'}{\del\lambda_i^{\,A}\over\del
x^{AB'}}+[a^A,\lambda_{iA}]\right)+
{\epsilon^{ijkl}\over4}\Phi_{ij}\left(\pi^{A'}{\del\phi_{kl}^{\,A}\over\del
x^{AA'}} + [a^A,\phi_{kl\,A}]\right)\\
\left.+{\epsilon^{ijkl}\over2}\Phi_{ij}\lambda_k^{\,A}\lambda_{lA}
+\left(b^A\delbar_0 a_A +\chi^{iA}\delbar_0\lambda_{iA}
+{\epsilon^{ijkl}\over8}\phi_{ij}^{\;A}\delbar_0\phi_{kl\,A}\right)\right\}
\label{huge}
\end{multline}
where we note that the expression inside the braces is weightless.
Since $S_2$ is independent of $\cA_A\bar e^A$ while the untilded
fields $b_A$ and $\chi^i_A$ appear here only linearly (in the last
line of the above formula), they play the role of Lagrange
multipliers. Integrating them out of the partition function
enforces $\bar e^0\delbar_0 a_A=\bar e^0\delbar_0\lambda_{i\,A}=0$
and so $a_A$ and $\lambda_{i\,A}$ must be holomorphic in $\pi$.
Since they have holomorphic weights +1 and 0 respectively, we find
\begin{equation}
a_A(x,\pi,\hat\pi) = A_{AA'}(x)\pi^{A'}
\quad\quad\hbox{and}\quad\quad
\lambda_{i\,A}(x,\pi,\hat\pi)= \Lambda_{i\,A}(x)
\label{ALambda}
\end{equation}
for some spacetime dependent fields $A_{AA'}$ and
$\Lambda_{i\,A}$. Similarly, because $\phi_{ij\,A}$ appears only
quadratically it may be eliminated\footnote{At the cost of a
field-independent determinant.} using its equation of motion
\begin{equation}
\delbar_0 \phi_{ij\,A} =\pi^{A'}\left({\del\Phi_{ij}\over\del
x^{AA'}} +[A_{AA'},\Phi_{ij}]\right) \label{phi}
\end{equation}
where we have used \ref{ALambda}. This implies
\begin{equation}
\phi_{ij\,A} ={1\over\pi\cdot\hat\pi}\hat\pi^{A'} D_{AA'}\Phi_{ij}
\end{equation}
where $D_{AA'}$ is the usual spacetime gauge covariant derivative
and, as in \ref{spcompts}, $\Phi$ depends only on spacetime
coordinates. Inserting our expressions for $a_A$, $\lambda_A$ and
$\phi_A$ into \ref{huge} now reduces the action to
\begin{multline}
S_1[\cA]={\im\over2\pi}\int{\Omega\wedge\bar\Omega\over(\pi\cdot\hat\pi)^4}
{\rm tr}\left\{{3\over2} B_{A'B'}
F_{C'D'}{\hat\pi^{A'}\hat\pi^{B'}\pi^{C'}\pi^{D'}\over(\pi\cdot\hat\pi)^2}+
2\tilde\Lambda^i_{\,A'}D^B_{\;\;B'}\Lambda_{iB}
{\hat\pi^{A'}\pi^{B'}\over\pi\cdot\hat\pi}\right.\\
\left.+{\epsilon^{ijkl}\over8}\Phi_{ij}D^A_{\;\;A'}D_{AB'}\Phi_{kl}
{\hat\pi^{A'}\pi^{B'}\over\pi\cdot\hat\pi}
+{\epsilon^{ijkl}\over2}\Phi_{ij}\Lambda_i^{\,A}\Lambda_{lA}\right\}
\label{n}
\end{multline}
where $F^{A'B'}$ is the selfdual part of the curvature of
$A_{AA'}$. None of the remaining fields depend on $\pi$ or
$\hat\pi$, so we can integrate out the fibres (see the appendix
for details). Doing so, one finds
\begin{equation}
S_1[\cA]=\int\rd^4x\;{\rm tr}\left\{{1\over2} B_{A'B'} F^{A'B'}
+\tilde\Lambda^i_{\,A'}D^{AA'}\Lambda_{iA} +
{\epsilon^{ijkl}\over16} D^A_{\;A'}\Phi_{ij}D_A^{\;A'}\Phi_{kl}+
{\epsilon^{ijkl}\over2}\Phi_{ij}\Lambda_k^{\,A}\Lambda_{lA}\right\}
\label{asdym}
\end{equation}
and, as is familiar from Witten's work~\cite{Witten:2003nn},
holomorphic Chern-Simons theory on $\bT^{3|4}$ thereby reproduces
the anti-selfdual interactions of $\cN=4$ SYM in an action first
discussed by Chalmers \& Siegel~\cite{Chalmers:1996rq}.

We must now find the contribution from $S_2$ and to do so, we must
vary the determinant.  The formula for the variation follows from
the prescription given earlier; we do not wish to give the full
theory here, but refer the reader to the discussion in section 3
of~\cite{Mason:2001}.  The device of renormalizing the metric on
the Quillen determinant line bundle was not used
in~\cite{Mason:2001}, but it simply has the effect of removing the
appearance of $\alpha^*$ from equation 3.3 of that paper
($\alpha^*$ is $\cA^*$ in our notation, with the $*$ denoting
complex conjugation with respect to a chosen Hermitian structure
on $E$). On restricting to gauge group SU($N$), we obtain
\begin{equation}
\delta\log\det\left(\left.\delbar_\cA\right|_L\right)= \int_L{\rm
tr}\,J\delta \cA
\end{equation}
where
\begin{equation}
 J(\pi_1)=\lim_{\pi_1\rightarrow \pi_2}
\left(G(\pi_1,\pi_2)-{1\over 2\pi\im}
 {I\over\pi_1\cdot\pi_2}\right) \pi_1\cdot\rd\pi_1
\end{equation}
in which $\pi_1$, $\pi_2$ are abbreviations for the homogeneous
coordinates $\pi_{1A'}$ on $L_1$ {\it etc.}, $G$ is the Greens
function for $\delbar_\cA$ on sections of $E$ of weight $-1$ over
$L$, $I$ is the identity matrix and $\pi$ denotes the usual ratio
of the circumference to the diameter of a circle.  Using the
relation
\begin{equation}
\delta G(\pi_1,\pi_2)=-\int_L G(\pi_1,\pi_3)\delta\cA
(x,\tilde\theta,\pi_3)G(\pi_3,\pi_2)\,\pi_3\cdot\rd\pi_3\, ,
\end{equation}
we can expand $S_2$ in powers of $\cA$ as
\begin{equation}
\log\det\left(\left.\delbar_\cA\right|_L\right)
= {\rm
tr}\left\{\ln\delbar_L+\sum_{r=1}^\infty {1\over
r}\left({-1\over2\pi\im}\right)^r\int
{\pi_1\cdot\rd\pi_1\over\pi_r\cdot\pi_1}\cA_1
{\pi_2\cdot\rd\pi_2\over\pi_1\cdot\pi_2}\cA_2
\cdots{\pi_r\cdot\rd\pi_r\over\pi_{r-1}\cdot\pi_r}\cA_r\right\}
\end{equation}
where $(1/2\pi\im)(1/\pi_i\cdot\pi_j)$ is the Green's function at
$\cA=0$\footnote{In the gauge \ref{spgauge}, the connection is
trivial along the fibres, so End$(E)$-valued fields may be
integrated over these fibres without worrying about parallel
propagation. We apologize for the proliferation of $\pi$s in our
Green's function, and hope the meaning is clear!}  for the
$\delbar_L$-operator on $L=\bP^1$. Each $\cA$ in this expansion is
restricted to lie on a copy of the fibre over the same point
$(x,\tilde\theta)$ in spacetime. In particular, they each depend
on the same $\tilde\theta^{A'i}$ so because
$\psi^i=\tilde\theta^{A'i}\pi_{A'}$ and $\cA_0\sim(\psi)^2$ in
this gauge, the series vanishes after the fourth term.
Furthermore, the measure $\rd\mu$ involves an integration
$\rd^8\tilde\theta$, so we only need keep the terms proportional
to $(\tilde\theta)^8$. Schematically then, the only relevant terms
are $B^2$, $\Phi\tilde\Lambda^2$, $\Phi^2 B$ and $\Phi^4$. In
fact, since $B_{A'B'}$ represents a selfdual 2-form on spacetime,
the $\Phi^2 B$ term may also be neglected since there is no way
for it to form a non-vanishing scalar once we integrate out the
$\bP^1$ fibre. The $B^2$ term is
\begin{equation}
-\kappa\int\rd\mu\,\frac12\left({3\over2\pi\im}\right)^2
\int\prod_{r=1}^2{K_r\over\pi_r\cdot\pi_{r+1}}{\rm tr}
\left\{{B_{A'B'}\hat\pi_1^{A'}\hat\pi_1^{B'}\over(\pi_1\cdot\hat\pi_1)^2}
{B_{C'D'}\hat\pi_2^{C'}\hat\pi_2^{D'}\over(\pi_2\cdot\hat\pi_2)^2}(\psi_1)^4(\psi_2)^4\right\},
\label{bsq}
\end{equation}
where we have defined the K\" ahler form
\begin{equation}
K={\pi\cdot\rd\pi\wedge\hat\pi\cdot\rd\hat\pi\over(\pi\cdot\hat\pi)^2}
\label{Kahler}
\end{equation}
on each copy of the $\bP^1$ fibre. The $\tilde\theta$ integrations
may be evaluated straightforwardly using Nair's lemma
\begin{equation}
\int\rd^8\tilde\theta\,\left.(\psi_1)^4(\psi_2)^4\right|_{L(x_-,\tilde\theta)}
=(\pi_1\cdot\pi_2)^4\, , \label{Nair}
\end{equation}
while the results in the appendix then allow us to integrate out
the fibres in  equation \ref{bsq}, yielding a contribution
$-{\kappa\over2}\int\rd^4x\;{\rm tr}2B_{A'B'}B^{A'B'}$ on
spacetime. To find the contributions from the
$\Phi\tilde\Lambda^2$ term
\begin{equation}
-\kappa\int\rd\mu\;
{2\over(2\pi\im)^3}\int\prod_{r=1}^3{K_r\over\pi_r\cdot\pi_{r+1}}{\rm
tr}\left\{\psi^i_1\psi^j_1\Phi_{ij}\epsilon_{klmn}{\psi^k_2\psi^l_2\psi^m_2\over3!}
{\tilde\Lambda^n_{A'}\hat\pi_2^{A'}\over\pi_2\cdot\hat\pi_2}
\epsilon_{pqrs}{\psi^p_3\psi^q_3\psi^r_3\over3!}
{\tilde\Lambda^s_{B'}\hat\pi_3^{B'}\over\pi_3\cdot\hat\pi_3}\right\}
\label{philamsq}
\end{equation}
and the $\Phi^4$ term
\begin{equation}
-\kappa\int\rd\mu\;\frac14{1\over(2\pi\im)^4}
\int\prod_{r=1}^4{K_r\over\pi_r\cdot\pi_{r+1}}
\psi^i_1\psi^j_1\psi^k_2\psi^l_2\psi^m_3\psi^n_3\psi^p_4\psi^q_4\;{1\over2^4}
{\rm tr}\left\{\Phi_{ij}\Phi_{kl}\Phi_{mn}\Phi_{pq}\right\}
\label{phi4}
\end{equation}
it is helpful to first integrate out the first copy of the fibre
from \ref{philamsq} and (say) the first and third copies from
\ref{phi4} using
\begin{equation}
\int K_1
{\pi_{1A'}\pi_{1B'}\over\pi_1\cdot\pi_2\,\pi_3\cdot\pi_1}\tilde\theta^{iA'}\tilde\theta^{jB'}
=-2\pi\im\;{\pi_{2A'}\pi_{3B'}+\pi_{3A'}\pi_{2B'}\over(\pi_2\cdot\pi_3)^2}\tilde\theta^{iA'}\tilde\theta^{jB'}
\,. \label{p1vol}
\end{equation}
These integrations reduce the $\tilde\theta$ dependence of
\ref{philamsq} and \ref{phi4} to the same form as in \ref{bsq};
integrating out these $\tilde\theta$s allows us to perform the
remaining fibre integrals as before. Combining all the terms, we
find that the $\log\det\delbar_\cA$ term provides a contribution
\begin{equation}
S_2[\cA]=-\kappa\int\rd^4x\,{\rm tr}\left\{\frac12 B_{A'B'}
B^{A'B'}+{1\over2}\Phi_{ij}
\tilde\Lambda^i_{\,A'}\tilde\Lambda^{j\,A'}+{1\over16}
\epsilon^{iklm}\epsilon^{jnpq}\Phi_{ij}\Phi_{kl}\Phi_{mn}\Phi_{pq}\right\}.
\label{sdym}
\end{equation}
Adding this to the Chern-Simons contribution in equation
\ref{asdym} gives the complete $\cN=4$ SYM action (up to the
topological invariant ${\rm c}_2(F)$); to put it in standard form
one integrates out $B_{A'B'}$, identifies $\kappa=g^2_{\rm YM}$
and rescales
$\tilde\Lambda_{A'}\rightarrow\tilde\Lambda_{A'}/\sqrt{g_{\rm
YM}}$, $\Lambda_A\rightarrow\sqrt{g_{\rm YM}}\Lambda_A$.

\subsection{The MHV formalism}

One of the pleasing features of the twistor action is that it
provides a simple way to understand the MHV diagram formalism of
Cachazo, Svr{\v c}ek \& Witten~\cite{Cachazo:2004kj}. Instead of
working in the gauge \ref{spgauge}, one picks an arbitrary spinor
$\eta^A$ and imposes the axial-like condition
$\eta^A\delbar_A\lrcorner\cA=0$. In this gauge, the $\cA^3$ vertex
of the Chern-Simons theory vanishes. However, we no longer have
the restriction that $\cA_0\sim(\psi)^2$, so the expansion
\begin{equation}
\log\det\left(\left.\delbar_\cA\right|_L\right) = {\rm
tr}\left\{\ln\delbar_L+\sum_{r=1}^\infty {1\over
r}\left({-1\over2\pi\im}\right)^r\int
{\pi_1\cdot\rd\pi_1\over\pi_r\cdot\pi_1}\cA_1
{\pi_2\cdot\rd\pi_2\over\pi_1\cdot\pi_2}\cA_2\cdots
{\pi_r\cdot\rd\pi_r\over\pi_{r-1}\cdot\pi_r}\cA_r\right\}
\label{MHV}
\end{equation}
in $S_2$ does not terminate. Focussing on the spin 1 sector, the
action contains an infinite series of vertices each of which is
quadratic in $B$ (so as to survive the $\tilde\theta$ integration)
and it is easy to see that these are exactly the MHV vertices.
Also, this gauge brings the substantial simplification that the
only non-vanishing components of on-shell fields are $\cA_0$. For
momentum eigenstates, the $\cA_0$ have delta function dependence
on $\pi_{A'}$ supported where $\pi_A'$ is proportional to the
corresponding spinor part of the spacetime momentum as
in~\cite{Witten:2003nn}. We have undertaken a study of
perturbation theory using this form of the action, and will
present our results in a companion paper~\cite{Boels:2006}.

\section{Theories with less supersymmetry}

Having dealt with the maximally supersymmetric gauge theory, let
us now study theories with $\cN=1$ \& 2 sets of spacetime
supercharges. Rather than work on weighted projective spaces, our
strategy here is to obtain (the SYM sector) of these theories by
breaking the U(4) $R$-symmetry of the $\cN=4$ theory. We will then
see how to couple these SYM theories to matter in an arbitrary
representation of the gauge group.

The $\cN=4$ theory possesses a U(4) $R$-symmetry which, in the
twistorial representation, arises from the freedom to rotate
$\psi$s into one another using the generators
$\psi^i\del/\del\psi^j$. To reach a theory with only $\cN=2$
supersymmetry one arbitrarily singles out two $\psi$ directions,
say $\psi^3$ and $\psi^4$, and demand that all fields depend on
them only via the combination $\psi^3\psi^4$ {\it i.e.} we require
invariance under the $R$-symmetry SU(2) in $(\psi^3,\psi^4)$. With
this restriction, the $\cN=4$ multiplet \ref{Aexp} becomes
\begin{equation}
\begin{aligned}
\cA &= a + \psi^a\lambda_a +
{1\over2}\epsilon_{ab}\psi^a\psi^b\phi
+\psi^3\psi^4\left(\tilde\phi +
\psi^a\chi_a+{1\over2}\epsilon_{ab}\psi^a\psi^b  b\right)\\
&=\cA^{(2)} + \psi^3\psi^4\cB^{(2)}
\end{aligned}\label{A2exp}
\end{equation}
where $a,b$ run from 1 to 2, and $\cA^{(2)}$ and $\cB^{(2)}$ have
the exact field content of an $\cN=2$ gauge multiplet and its CPT
conjugate. Upon integrating out $\psi^3\psi^4$, the action
$S_1+S_2$ becomes (dropping the wedges)
\begin{multline}
S_{\rm
gauge}[\cA^{(2)},\cB^{(2)}]={\im\over2\pi}\int\Omega\rd^2\psi\,
{\rm tr}\,\cB^{(2)}\cF^{(2)}\\
+{\kappa\over8\pi^2}\int\rd^4x\rd^4\tilde\theta
(\pi_1\cdot\pi_2)^2{\rm tr}\left\{(\delbar +
\cA^{(2)})^{-1}_{21}\cB^{(2)}_1\,(\delbar+\cA^{(2)})_{12}^{-1}\cB^{(2)}_2\right\}
\label{2act}
\end{multline}
where $\cF^{(2)}=\delbar \cA^{(2)}+ [\cA^{(2)},\cA^{(2)}]$ is the
curvature of $\cA^{(2)}$. The definition \ref{A2exp} implies that
$\cB^{(2)}$ has holomorphic weight -2 so that this action is
well-defined on the projective space. The integrand in the second
term of this action is understood to be restricted to copies of
the $\bP^1$ fibres over $(x_-,\tilde\theta)$ as in section 3. The
subscripts on the $\cB$ fields and the Green's functions in this
term label copies of the fibres, while $(\delbar+\cA)^{-1}_{ij}$
is understood to involve an integral over fibre $j$.   Keeping
only the appropriate components of the fields, it is
straightforward to verify that \ref{2act} reproduces the standard
$\cN=2$ spacetime SYM action (up to a non-perturbative term) when
the gauge \ref{spgauge} is imposed.

Notice that this method of restricting the dependence of $\cA$ on
the fermionic coordinates is similar to, but distinct from,
working on a weighted projective superspace. Although
$\psi^3\psi^4$ is a nilpotent object of weight 2, it is bosonic
and we would not have obtained the above action from a string
theory on the weighted Calabi-Yau supermanifold
$\bW\bP^{3|3}(1,1,1,1|1,1,2)$. It is also interesting to consider
the effect of the scaling $\psi\mapsto r\psi$. The action
\ref{2act} is invariant under the U(1) (really, $\bC^*$) part of
the remaining U(2) $R$-symmetry if we shift the charge of
$\cB^{(2)}$ so that $\psi^a\mapsto r\psi^a$ induces
$\cB^{(2)}\mapsto r^2\cB^{(2)}$. The component fields
$\{a,\lambda_a,\phi\}$ and $\{\tilde\phi,\chi_a,b\}$ then have
charges $\{0,-1,-2\}$ and $\{2,1,0\}$ respectively, exactly the
grading of these fields that is familiar from Donaldson-Witten
theory, for example.

Similarly, to obtain $\cN=1$ SYM one demands that $\cA$ depends on
$\psi^2$, $\psi^3$ and $\psi^4$ only through the combination
$\psi^2\psi^3\psi^4$ so that, calling $\psi^1=\psi$,
\begin{equation}
\begin{aligned}
\cA&= a + \psi\lambda - \psi^2\psi^3\psi^4(\chi + \psi b)\\
&= \cA^{(1)} -\psi^2\psi^3\psi^4\cB^{(1)}
\end{aligned}\label{A1exp}
\end{equation}
with $\cA^{(1)}$ and $\cB^{(1)}$ containing exactly the field
content of an $\cN=1$ gauge multiplet and its CPT conjugate. The
constraint that $\psi^2\psi^3\psi^4$ always appear together leaves
no room for $\phi$, and the action is simply
\begin{multline}
S_{\rm gauge}[\cA^{(1)},\cB^{(1)}]
={\im\over2\pi}\int\Omega\rd\psi\, {\rm tr}\,\cB^{(1)}\cF^{(1)}\\
+{\kappa\over8\pi^2}\int\!\rd^4x\rd^2\tilde\theta
(\pi_1\cdot\pi_2)^3{\rm
tr}\left\{(\delbar+\cA^{(1)})_{21}^{-1}\cB_1^{(1)}(\delbar+\cA^{(1)})_{12}^{-1}\cB^{(1)}_2\right\}.
\label{3act}
\end{multline}
In this case, in spacetime gauge only the $B^2$ term survives from
$S_2$, since all others involved $\phi$. Again, it is
straightforward to check that this gauge choice yields exactly the
usual $\cN=1$ action, and that the residual $r$-scaling is just
the usual U(1) $R$-symmetry.

\subsection{Matter multiplets}

In theories with $\cN<4$ supersymmetries, additional multiplets
are possible. At $\cN=2$ there is a hypermultiplet consisting of
fields with helicities $(-\frac12^1,0^2,+\frac12^1)$ together with
its CPT conjugate, where the superscripts denote multiplicity. At
$\cN=1$ we have a chiral multiplet whose component fields have
helicities $(-\frac12^1,0^1)$ together with its antichiral CPT
conjugate. These multiplets were first constructed in twistor
superspaces by Ferber~\cite{Ferber:1977qx} and take the forms
\begin{equation}
\cN=2 {\hbox{ hyper}}\left\{
\begin{aligned}
\cH &= \rho+\psi^ah_a +{\epsilon_{ab}\over2}\psi^a\psi^b\tilde\mu\\
\tilde\cH &=\mu + \psi^a\tilde h_a + {\epsilon_{ab}\over2}\psi^a\psi^b\tilde\rho
\end{aligned}\right.
\label{hyper}
\end{equation}
where $\cH$ and $\tilde\cH$ are each fermionic and have weight $-1$, and
\begin{equation}
\cN=1 {\hbox{ chiral}}\left\{
\begin{aligned}
\cC &= \nu + \psi m\\
\tilde\cC &= \tilde m + \psi\tilde\nu
\end{aligned}\right.
\label{chiral}
\end{equation}
where $\cC$ is fermionic and of weight $-1$, while $\tilde\cC$ is
bosonic and of weight $-2$; all the above fields are (0,1)-forms.
The matter fields may take values in arbitrary representations $R$
of the gauge group. Their actions take similar forms, for example
\begin{multline}
S_{\rm hyp}[\cH,\tilde\cH,\cA^{(2)}] = \int\Omega\,\rd^2\psi\,
{\rm tr}\,\left\{\tilde\cH\,\delbar_{\cA^{(2)}}\cH\right\}\\
+2\kappa\int\rd^4x\,\rd^4\tilde\theta\;{\rm tr}\left\{
\left(\delbar_{\cA^{(2)}}\right)^{-1}_{31}\cH_1
\left(\delbar_{\cA^{(2)}}\right)^{-1}_{12}\tilde\cH_2
\left(\delbar_{\cA^{(2)}}\right)^{-1}_{23}\cB^{(2)}_3\,
\pi_1\cdot\pi_3\,\pi_2\cdot\pi_3\right\}\\
-{3\kappa\over2}\int\rd^4x\,\rd^4\tilde\theta\;{\rm tr}\left\{
\left(\delbar_{\cA^{(2)}}\right)^{-1}_{41}\cH_1
\left(\delbar_{\cA^{(2)}}\right)^{-1}_{12}\tilde\cH_2
\left(\delbar_{\cA^{(2)}}\right)^{-1}_{23}\cH_3
\left(\delbar_{\cA^{(2)}}\right)^{-1}_{34}\tilde\cH_4\;
\pi_1\cdot\pi_3\,\pi_2\cdot\pi_4\right\} \label{hypact}
\end{multline}
for a hypermultiplet in the fundamental representation and
\begin{multline}
S_{\rm ch}[\cC,\tilde\cC,\cA^{(1)}] = \int\Omega\,\rd\psi\, {\rm
tr}\,\left\{\tilde\cC\,\delbar_{\cA^{(1)}}\cC\right\}\\
+2\kappa\int\rd^4x\,\rd^2\tilde\theta\;{\rm tr}\left\{
\left(\delbar_{\cA^{(1)}}\right)^{-1}_{31} C_1
\left(\delbar_{\cA^{(1)}}\right)^{-1}_{12} \tilde C_2
\left(\delbar_{\cA^{(1)}}\right)^{-1}_{23}\cB^{(1)}_3 \,
\pi_1\cdot\pi_2(\pi_2\cdot\pi_3)^2\right\}\\
-{3\kappa\over2}\int\rd^4x\,\rd^2\tilde\theta\;{\rm tr}\left\{
\left(\delbar_{\cA^{(1)}}\right)^{-1}_{41} C_1
\left(\delbar_{\cA^{(1)}}\right)^{-1}_{12} \tilde C_2
\left(\delbar_{\cA^{(1)}}\right)^{-1}_{23} C_3
\left(\delbar_{\cA^{(1)}}\right)^{-1}_{34}\tilde C_4\;
(\pi_1\cdot\pi_3)^2(\pi_2\cdot\pi_4)^2 \right\} \label{chact}
\end{multline}
for a fundamental chiral multiplet, where the traces and
$\delbar_\cA$-operators are in the fundamental representation. The
actions are well-defined on the projective superspaces, with the
weights of the measures being balanced by those of the fields.
Again, the subscripts label the copy of the fibre on which the
relevant field is to be evaluated, and the operators
$\left(\delbar_{\cA}\right)^{-1}_{ij}$ involve an integral over
the $j^{\rm th}$ fibre. These actions may be obtained by symmetry
reduction, using the decomposition of the $\cN=4$ gauge multiplet
into $\cN=2$ gauge and hyper-multiplets, or $\cN=1$ gauge and
chiral multiplets, and then changing the representation (and
number) of matter multiplets. In fact all these matter couplings
can be obtained by an appropriate symmetry reduction from some
large gauge group and the $\int \rd^4x\rd^8\theta$  expressions in
\ref{hypact} and \ref{chact} may be understood in that context as
additional contributions from a `log det' term.

Since the matter fields are all (0,1)-forms, there is an
additional symmetry that may be surprising from the spacetime
perspective. For example, when $\cC$ is in the fundamental
representation while $\tilde\cC$ is in the antifundamental, then
the complete $\cN=1$ action $S_{\rm gauge}+S_{\rm ch}$ is
invariant under the usual gauge transformations
\begin{equation}
\begin{aligned}
\delbar+\cA^{(1)} &\rightarrow g(\delbar+\cA^{(1)})g^{-1}\hspace{1cm} &\cC &\rightarrow g\,\cC\\
\cB^{(1)}&\rightarrow g\cB^{(1)}g^{-1} &\tilde\cC &\rightarrow \tilde\cC g^{-1},
\end{aligned}\label{matgauge}
\end{equation}
but it is also invariant under the transformations
\begin{equation}
\cC\rightarrow \cC + \delbar_{\cA^{(1)}}M\quad\quad
\tilde\cC\rightarrow \tilde\cC + \delbar_{\cA^{(1)}}\tilde
M\quad\quad \cB^{(1)}\rightarrow \cB^{(1)} + \cC\tilde M -
M\tilde\cC \label{extra}
\end{equation}
where $M\in\Gamma_{\bT^{3|1}}(E(-1))$ is a fermion
and $\tilde M\in\Gamma_{\bT^{3|1}}(E^*(-2))$ is a
boson. The fact that the matter fields are only defined up to
exact forms is a direct consequence of the fact that physical
information is encoded in cohomology on twistor space. To evaluate
any path integral involving these matter fields, this additional
symmetry needs to be fixed. In particular, requiring
$\delbar_L^*\cC=0$ and $\delbar_L^*\tilde\cC=0$ on each fibre $L$
allows one to reduce the theory to (the kinetic and D-term parts
of) the usual spacetime action, in exactly the same way as was
done in section 3. Here, no residual freedom remains once these
conditions are imposed because the fields $M$ and $\tilde M$ each
have negative weight, but $H^0(\bP^1,\cO(n))=0$ for $n<0$.

\section{Discussion}

We have studied actions for twistorial gauge theories, showing how
they are related to the standard spacetime and MHV formalisms. A
detailed investigation of perturbation theory using this action
will be presented in a companion paper~\cite{Boels:2006}. However,
the demonstration that $S_1+S_2$ is perturbatively equivalent to
$\cN=4$ SYM in spacetime at the level of the partition function
makes it clear that conformal supergravity does not appear in our
treatment.

It is instructive to contrast our picture with the original
twistor-string proposal. Scattering amplitudes between states with
wavefunctions $\cA_1,\ldots,\cA_n$ may be obtained in any quantum
field theory by varying the generating functional
\begin{equation}
\e^{F[\cA_{\rm cl}]}=\int_{\cA\to\cA_{\rm cl}}\!\!\!\!\!\!\!\!
{\rm D}\cA\;\e^{-S[\cA]} \label{genfun}
\end{equation}
with respect to $\cA_{\rm cl}$ in the directions
$\cA_1,\ldots,\cA_n$ and evaluating at $\cA_{\rm cl}=0$, where the
path integral in \ref{genfun} is taken over field configurations
that approach $\cA_{\rm cl}$ asymptotically. Witten
conjectured~\cite{WittenOxford} that the free energy for
twistor-strings could be evaluated as
\begin{equation}
\e^{F[\cA_{\rm cl}]} = \sum_{g=0,d=1}^\infty\kappa^d\int_{\cM^{\rm
conn}_{g,d}}\rd\mu_{g,d}\; \det\left(\left.\delbar_{\cA_{\rm
cl}}\right|_{C'}\right) \label{tsconj}
\end{equation}
where $\cM_{g,d}^{\rm conn}$ is a contour in the moduli space of
connected, genus $g$ degree $d$ curves in $\bT^{3|4}$,
$\rd\mu_{g,d}$ is some measure on $\cM_{g,d}^{\rm conn}$ and
$C'\in\cM_{g,d}^{\rm conn}$. In the disconnected prescription,
this conjecture may be recast as
\begin{equation}
\e^{F[\cA_{\rm cl}]} = \int_{\cA\to\cA_{\rm
cl}}\!\!\!\!\!\!\!\!{\rm
D}\cA\;\e^{-S_1[\cA]}\left\{\sum_{d=1}^\infty\kappa^d\int_{\cM_d}\rd\mu_d\;
\det\left(\left.\delbar_{\cA}\right|_{C}\right)\right\}
\label{lionel}
\end{equation}
as was argued in~\cite{Mason:2005zm}. Here, $S_1[\cA]$ is the
holomorphic Chern-Simons action, $\cM_d$ is a contour in the
moduli space of maximally disconnected, genus zero degree $d$
curves in $\bT^{3|4}$ and $C\in\cM_d$. In this formula, the effect
of the functional integral is to introduce Chern-Simons
propagators connecting the different disconnected components of
the curves together. In \cite{Gukov:2004ei} Gukov, Motl \& Neitzke
argued that these two formulations of twistor-string theory could
be shown to be equivalent by deforming the contour in the space of
curves through regions in which components of the disconnected
curves come together in such a way as to eliminate the
Chern-Simons propagators and connect the curves.

We can obtain an effective action that would lead to Witten's
conjecture as follows. First, choose the contour $\cM_d$ to be
$(\bE^{4|8})^d/{\rm Sym}_d$, the set of unordered $d$-tuples
$(x_i,\tilde\theta_i), i=1,\ldots, d$ of (possibly coincident)
points in $\bE^{4|8}$, so that the degree $d$ curve $C$ is the
union $C=\cup_{i=1}^d L(x_i,\tilde\theta_i)$ of $d$ lines. On a
disconnected curve, the determinant factorizes to give
\begin{equation}
\det\left(\left.\delbar_\cA\right|_{C}\right)=\prod_{i=1}^d
\det\left(\left.\delbar_\cA\right|_{L(x_i,\tilde\theta_i)}\right)\,.
\label{factorize}
\end{equation}
Similarly, the measure $\rd\mu_d$ on $\cM_d$ can be written
$\rd\mu_d=\prod_{i=1}^d \rd^4 x_i\rd^8\tilde\theta_i$. Putting
this together, we find that the conjecture implies
\begin{eqnarray}
\e^{F[\cA_{\rm cl}]}&=&\int{\rm D}\cA\; \e^{-S_1[\cA]}\left\{
\sum_{d=1}^\infty{\kappa^d\over
d!}\int_{(\bE^{4|8})^d}\prod_{i=1}^d \rd^4x_i\rd^8\tilde\theta_i\;
\det\left.\delbar_\cA\right|_{L(x_i,\tilde\theta_i)}\right\}\nonumber\\
&=& \int{\rm D}\cA\; \e^{-S_1[\cA]}\left\{\sum_{d=1}^\infty
{1\over
d!}\left(\kappa\int_{\bE^{4|8}}\rd^4x\rd^8\tilde\theta\;\det
\left.\delbar_\cA\right|_{L(x,\tilde\theta)}\right)^d\right\}\\
&=& \int{\rm D}\cA\;\exp\left(-S_1[\cA]
+\kappa\int_{\bE^{4|8}}\rd^4x\rd^8\tilde\theta\;\det\left.\delbar_\cA\right|_{L(x,\tilde\theta)}\right),
\nonumber\label{fakeact}
\end{eqnarray}
where the $1/d!$ factors take account of the ${\rm Sym}_d$ in the
definition of $\cM_d$. Thus, instead of the
$S_2[\cA]=-\kappa\int\log\det\delbar_\cA$ in our theory, the
twistor-string inspired conjecture would seem to require the
different action $\widetilde S_2[\cA]=-\kappa\int\det\delbar_\cA$.
However, expanding $\widetilde S_2[\cA]$ in $\cA$ shows that this
latter form contains spurious multi-trace terms, so these are
present in the original twistor-string proposal even at the level
of the action. Moreover, $\widetilde S_2[\cA]$ is not gauge
invariant because of the behaviour of the determinant discussed in
section 3. Restoration of gauge invariance can only be achieved at
the cost of coupling to the closed string sector. As we have
emphasized, the action of section 3 possesses neither of these
unwelcome features.

It is of course of great interest to see whether the action of
section 3 can be given a string interpretation and a `connected
prescription' found. While we do not yet have a full understanding
of this, the following remarks may be of interest. The natural
observables of real Chern-Simons theory on a three manifold (say
$S^3$) are the Wilson loops $W_{R}(\gamma)={\rm tr}_R {\rm
P}\exp\oint_\gamma A$ depending on some representation $R$ of the
gauge group. The correlation function~\cite{Witten:1988hf}
\begin{equation}
\langle \prod W_{R_i}(\gamma_i)\rangle = \int\rD
A\,\exp\left({1\over4\pi}\int_{S^3} {\rm tr}\left\{A\rd
A+{2\over3}A^3\right\}\right)\;\prod W_{R_i}(\gamma_i)
\label{Wilson}
\end{equation}
computes link invariants of the curves $\gamma_i\subset S^3$
depending on representations $R_i$. The Chern-Simons theory on
$S^3$ may be interpreted as the open string field theory of the
A-model on $T^*S^3$~\cite{Witten:1992fb} and the Wilson loops
themselves find a stringy interpretation in terms of Lagrangian
branes $L_i\subset T^*S^3$ with $L_i \cap S^3=\gamma_i$. The field
theory on the worldvolume of a single such brane wrapping $L_i$
contains a complex scalar in an $N$-dimensional representation $R$
of the gauge group of the Chern-Simons theory on the $S^3$.
Integrating out this scalar produces
$\det^{(R)}\rd_A|_{\gamma_i}$. This determinant may be related to
the holonomy around $\gamma_i$ by the formula
\begin{equation}
{\det}^{(R)} \rd_A|_\gamma = \det\left(1-\left({\rm P}\exp\oint_\gamma A\right)_R\right)
\label{segal}
\end{equation}
which follows from $\zeta$-function regularization (see {\it
e.g.}~\cite{Segal}). Hence, the Chern-Simons expectation value
\begin{equation}
\langle\det\rd_A|_\gamma\rangle =\int{\rm D}A\; \exp\left(-S_{\rm
CS}[A]+ \log\det\left(1-{\rm P}\exp\oint_\gamma A\right)\right)
\label{unknots}
\end{equation}
may be viewed as a generating functional for all the observables
associated to the knot $\gamma$~\cite{Ooguri:1999bv} upon
expanding in powers of the holonomy. Notice that the effective
action here is $\log\det{\rm d_A}$.

The partition function we presented in section 3 may be formally
understood to arise as the expectation value of an infinite
product of determinants in the holomorphic Chern-Simons theory
\begin{equation}
\int{\rm D}\cA\;\e^{-S_1[\cA]}\prod_{(x_-,\tilde\theta)}\left.\det
\delbar_\cA\right|_{L(x_-,\tilde\theta)} = \int{\rm
D}\cA\;\e^{-S_1[\cA]}
\exp\left(\int_{\bE^{4|8}_-}\rd\mu\,\left.\log\det
\delbar_\cA\right|_{L(x_-,\tilde\theta)}\right), \label{holwils}
\end{equation}
so it is tempting to interpret this as the generating functional
for all observables associated to all possible degree 1
holomorphic curves in $\bT^{3|4}$. In searching for a string
interpretation, we would like to find objects which support only
certain types of amplitude, graded by $d$ as in \ref{scaling}. To
this end, one might seek an analogue of knot invariants for
holomorphic curves. Holomorphic linking has been far less studied
than real linking (see~\cite{Frenkel:2005qk,Thomas} for the
Abelian case), but it may be exactly what is needed here
(see~\cite{Atiyah:1981ey,PenroseTN} for earlier discussions of
holomorphic linking in twistor space). In order to study link
invariants in the real category, one needs to supply framings both
of the underlying 3-manifold $M$ and of the knots $\gamma_i\in M$,
and from the Chern-Simons or A-model point of view, framings arise
via a coupling to the gravitational or closed string sector.
However, the expectation value \ref{Wilson} depends on the choice
of framing only through a simple phase, and it is perfectly
possible to make sense of link invariants nonetheless. One might
hope that the closed string sector of the twistor-string is no
more harmful.

In our view, it seems as though the ingredients of twistor-string
theory are correct -- perhaps only the recipe needs adjusting. We
hope that the considerations we have presented in this paper will
help to illuminate the story further.

\vspace{2cm}

\noindent{\large\bf Acknowledgments}

The authors would like to thank Philip Candelas and Wen Jiang for
discussions. DS acknowledges the support of EPSRC (contract number
GR/S07841/01). The work of LM and RB is supported by the European
Community through the FP6 Marie Curie RTN {\it ENIGMA} (contract
number MRTN-CT-2004-5652).  LM is also supported by a Royal
Society Leverhulme Trust Senior Research Fellowship.

\newpage
\appendix

\section{Integrating over the fibres}

In showing that our twistor actions reduce to spactime ones, it is
necessary to integrate over the $\bP^1$ fibres of $\bT\rightarrow
\bE$. Specifically, in equation \ref{n} we needed to integrate
expressions of the generic type
\begin{equation}
\int {\Omega\wedge\bar\Omega\over(\pi\cdot\hat\pi)^4}\,
S_{A'B'\ldots}T_{C'D'\ldots}{\pi^{A'}\pi^{B'}\cdots\hat\pi^{C'}\hat\pi^{D'}\cdots\over(\pi\cdot\hat\pi)^n}
\label{a1}
\end{equation}
where $S,T$ are spacetime dependent tensors with $n$ indices each;
in fact, in all the cases that arise in this paper, $S\in{\rm
Sym}_n\, \bS^+$. We start by noting that this integral is
well-defined on the projective twistor space, and hence on each
$\bP^1$ fibre. From \ref{frame}-\ref{emeas} we have
\begin{eqnarray}
{\Omega\wedge\bar\Omega\over(\pi\cdot\hat\pi)^4} &=&
{\pi\cdot\rd\pi\wedge\hat\pi\cdot\rd\hat\pi\over(\pi\cdot\hat\pi)^4}\wedge
\left(\rd x^{AA'}\wedge\rd x^{BB'}\wedge\rd x^{CC'}\wedge\rd
x^{DD'}\right)
\pi_{A'}\pi_{B'}\hat\pi_{C'}\hat\pi_{D'}\epsilon_{AB}\epsilon_{CD}\nonumber\\
&=&\rd^4x
{\pi\cdot\rd\pi\wedge\hat\pi\cdot\rd\hat\pi\over(\pi\cdot\hat\pi)^2}.
\label{a2}
\end{eqnarray}
where we have used
$\epsilon^{abcd}=\epsilon^{AD}\epsilon^{BC}\epsilon^{A'C'}\epsilon^{B'D'}
-\epsilon^{AC}\epsilon^{BD}\epsilon^{A'D'}\epsilon^{B'C'}$ and we
remind the reader that our $\sigma$-matrices are normalized so
that $\sigma^2=1$. Hence \ref{a1} becomes
\begin{equation}
\int_\bE\rd^4x
\int_{\bP^1}{\pi\cdot\rd\pi\wedge\hat\pi\cdot\rd\hat\pi\over(\pi\cdot\hat\pi)^{n+2}}
S_{A'B'\ldots}T_{C'D'\ldots}\pi^{A'}\pi^{B'}\cdots\hat\pi^{C'}\hat\pi^{D'}\cdots
\label{a3}
\end{equation}
which is the also form that arises  when reducing $S_2[\cA]$ to
spacetime. An object $S:=S_{A'B'\ldots}\pi^{A'}\pi^{B'}\cdots$
with $n$ $\pi$s is annihilated by the $\delbar$-operator on the
$\bP^1$ and hence is an element of $H^0(\bP^1,\cO(n))$. On the
other hand,
\begin{equation}
T:=T_{C'D'\ldots}{\hat\pi^{C'}\hat\pi^{D'}\cdots\over(\pi\cdot\hat\pi)^{n+2}}\hat\pi\cdot\rd\hat\pi
\label{a4}
\end{equation}
is also $\delbar$-closed (since ${\rm dim}_\bC\,\bP^1=1$) and is
in fact harmonic~\cite{Woodhouse:1985id} (indeed, we used this in
the text to solve the gauge condition \ref{spgauge}). Thus it
represents an element of $H^1(\bP^1,\cO(-n-2))$. Serre duality
asserts that
\begin{equation}
H^1(\bP^1,\cO(-n-2)\simeq H^0(\bP^1,\Omega^1(\cO(n+2)))
\label{a5}
\end{equation}
and in our case the duality pairing is given by
\begin{equation}
{1\over2\pi\im}\int_{\bP^1}\left(\pi\cdot\rd\pi\; S\right)\wedge T
= -{1\over n+1}S_{A'B'\ldots}T^{A'B'\ldots}
\label{a6}
\end{equation}
which is straightforward to check explicitly by working in local
coordinates on the $\bP^1$. Hence we find
\begin{equation}
\int_\bT{\Omega\wedge\bar\Omega\over(\pi\cdot\hat\pi)^4}\,
S_{A'B'\ldots}T_{C'D'\ldots}{\pi^{A'}\pi^{B'}\cdots\hat\pi^{C'}\hat\pi^{D'}\cdots\over(\pi\cdot\hat\pi)^n}
=-{2\pi\im\over(n+1)}\int_\bE \rd^4x\;S_{A'B'\ldots}T^{A'B'\ldots}
\label{a7}
\end{equation}
which was used in section 3.

\end{document}